# 3D MC. II. X-Ray Echoes Reveal a Clumpy Molecular Cloud in the CMZ

Danya Alboslani[1] , Cara Battersby[1] , Samantha W. Brunker[1] , Maïca Clavel[2] , Dani Lipman[1] , and Daniel L. Walker[1,3]
[1] University of Connecticut, Department of Physics, 196A Auditorium Road, Unit 3046, Storrs, CT 06269, USA
[2] Univ. Grenoble Alpes, CNRS, IPAG, F-38000 Grenoble, France
[3] ALMA Regional Centre Node, Jodrell Bank Centre for Astrophysics, The University of Manchester, Manchester M13 9PL, UK

## Abstract

X-ray observations collected over the past decades have revealed a strongly variable X-ray signal within the Milky Way's Galactic center, interpreted as X-ray echoes from its supermassive black hole, Sgr A*. These echoes are traced by the strong Fe Kα fluorescent line at 6.4 keV, the intensity of which is proportional to the density of the illuminated molecular gas. Over time, the echo scans through molecular clouds (MCs) in our Galactic center, revealing their 3D structure and highlighting their densest parts. While previous studies have utilized spectral line Doppler shifts along with kinematic models to constrain the geometry of the Central Molecular Zone (CMZ) or to study the structure of individual clouds, these methods have limitations, particularly in the turbulent region of the CMZ. We use archival Chandra X-ray data to construct one of the first 3D representations of one prominent MC, the Stone cloud, located at ($\ell = 0°.068$, $b = -0°.076$) at a distance of ~20 pc from Sgr A* in projection. Using the Chandra X-ray Observatory, we followed the X-ray echo in this cloud from 2008 to 2017. We combine these data with 1.3 mm dust continuum emission observed with the Submillimeter Array (SMA) and the Herschel Space Observatory to reconstruct the 3D structure of the cloud and estimate the column densities for each year's observed slice. The analysis of the X-ray echoes, along with velocities from SMA molecular line data, indicates that the structure of the Stone cloud can be described as a very diffuse background with multiple dense clumps throughout.

*Unified Astronomy Thesaurus concepts:* Molecular clouds (1072); Galactic center (565); X-ray astronomy (1810)

## 1. Introduction

The inner ~300 pc of the Milky Way's Galactic center, the Central Molecular Zone (CMZ), is often characterized as an extreme environment with molecular hydrogen contained in molecular clouds (MCs) with gas densities exceeding $10^4$ cm$^{-3}$ (e.g., R. Guesten & C. Henkel 1983; E. A. C. Mills et al. 2018), gas temperatures of 50–100 K (A. Ginsburg et al. 2016; N. Krieger et al. 2017), intense magnetic fields of 10–1000 μG (e.g., N. L. Chapman et al. 2011; N. O. Butterfield et al. 2024; T. Pillai et al. 2015), and high velocity dispersions of >100 km s$^{-1}$ (e.g., M. C. Sormani et al. 2019). Furthermore, the Galactic center of the Milky Way varies from the Galactic disk environment, where temperatures, turbulence, densities, and magnetic field strengths are about an order of magnitude lower (J. D. Henshaw et al. 2023). While the conditions in the Galactic center are dissimilar to those of the Galactic disk, they are congruent with those found in starburst (R. M. Crocker 2012), ultraluminous infrared, and high-redshift galaxies (J. M. D. Kruijssen & S. N. Longmore 2013), which are known for their high star formation (e.g., N. Seymour et al. 2008; S. Garcìa-Burillo et al. 2012). Despite the CMZ having a similar environment (M. D. Kruijssen & S. N. Longmore 2013), it produces stars an order of magnitude below what it should for the amount of dense gas present (K. Immer et al. 2012; S. N. Longmore et al. 2013), generating interest in studying the molecular gas properties and dynamics in the CMZ in recent years.

The CMZ contains roughly $(2–6) \times 10^7 M_\odot$ (G. Dahmen et al. 1998; K. Ferrière et al. 2007; C. Battersby et al. 2024b), corresponding to 3%–10% of the total molecular gas in the

Galaxy (J. Roman-Duval et al. 2016). This molecular gas is concentrated in MCs of different sizes and densities that can form clumps and cores, within which stars can form under the right conditions. The cool temperature of molecular gas and lack of a permanent electric dipole moment of the most abundant molecule, H$_2$, make it difficult to study MCs using H$_2$. Instead, examples of molecular lines used to map MCs include CO (e.g., T. Oka et al. 1996, 1998; T. M. Bania 1977), NH$_3$ (e.g., C. R. Purcell et al. 2012), and H$_2$CO (e.g., A. J. Walsh et al. 2011). Other molecules such as HCN or SiO can be used to trace the densest parts of these clouds, which can be shocked (J. Martìn-Pintado et al. 1997). Furthermore, radio and submillimeter telescopes have also helped to map the distribution of the densities of these MCs in projection (e.g., Y. Sofue & T. Handa 1984; G. Novak et al. 2003; C. Battersby et al. 2020).

Previous studies show that the molecular gas flows along the galactic bar and into the inner CMZ (N. M. McClure-Griffiths et al. 2012; M. C. Sormani & A. T. Barnes 2019; H. P. Hatchfield et al. 2021; Y. Su et al. 2024) at a distance of approximately 100 pc with a semimajor axis perpendicular to the bar (J. Binney et al. 1991; M. C. Sormani et al. 2015). Most of the gas in the CMZ, as well as the dense MCs, lies on an approximately elliptical orbit, called the Milky Way's inner $x_2$ orbit, surrounding Sgr A*. The exact orbital model is widely debated, but current models include a closed ellipse, an open stream, and two spiral arms (Y. Sofue 1995; S. Molinari et al. 2011; J. M. D. Kruijssen et al. 2015). The most recent literature shows that the best-fit orbital model is an improved version of the J. M. D. Kruijssen et al. (2015) ellipse model (D. Lipman et al. 2024; D. L. Walker et al. 2024). Only by uncovering the 3D distribution of MCs and gas in the CMZ can we understand the connection between gas inflow, star formation, black hole feeding, and outflow.

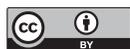

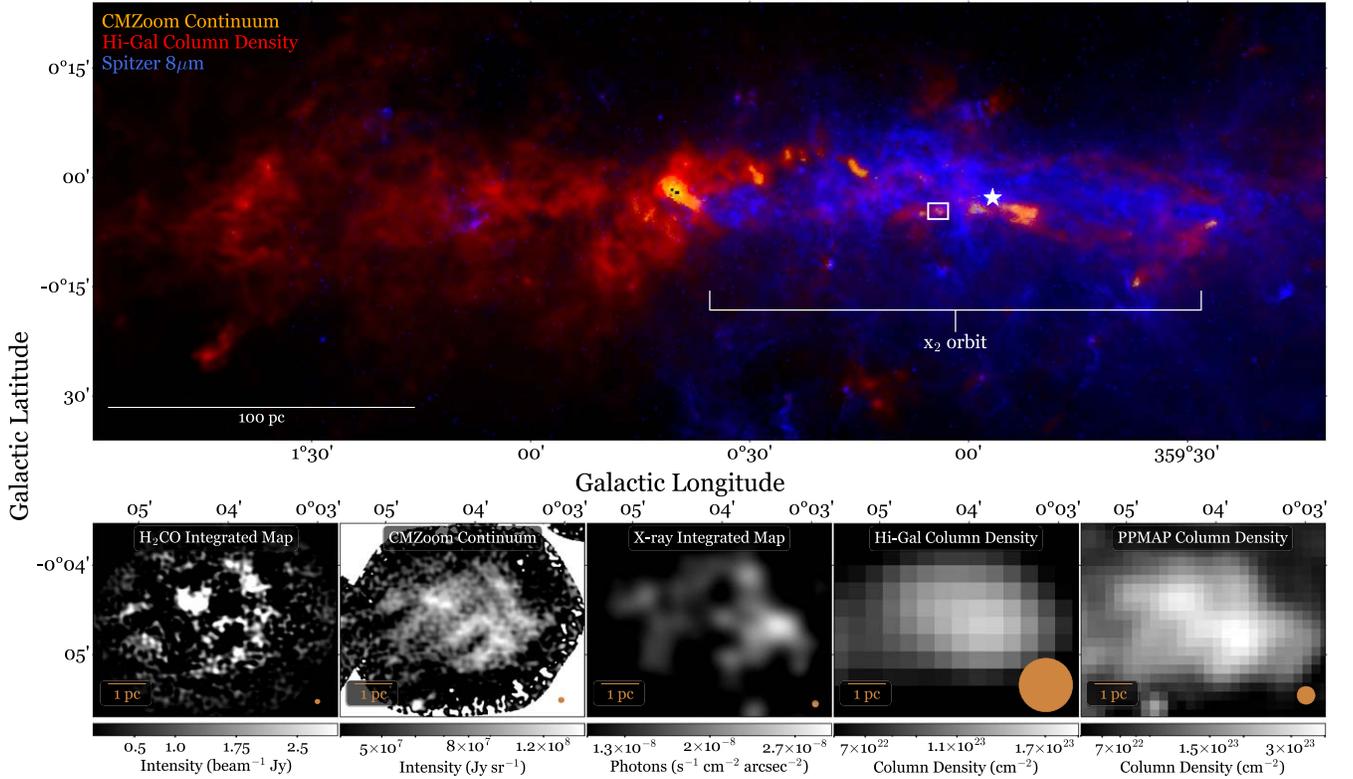

**Figure 1.** Top: RGB image of the CMZ (red: Hi-GAL column density; blue: Spitzer 8 μm; orange: CMZoom continuum). The white star represents the position of Sgr A*, while the white rectangle shows the position of the Stone MC. The bottom five panels represent the data used for analysis in this paper. From left to right: $H_2CO$ integrated map observed by the Submillimeter Array, Submillimeter Array CMZoom continuum, Chandra X-ray integrated emission from 2008 to 2017, Herschel Hi-GAL column density, and PPMAP column density. Scale bars are given in orange in the lower left corner of each panel, while the beam sizes are shown in the lower right corner as orange circles.

The individual 3D structure of MCs is hard to accurately model in the CMZ (C. N. Beaumont et al. 2013). This is partly due to the fact that extreme extinction in the CMZ ($A_V \gtrsim 30$ mag, $A_K \gtrsim 2.5$ mag; S. Nishiyama et al. 2008; R. Schödel et al. 2010), along with crowding, hampers observations and extinction methods of the CMZ. Previous studies have probed the 3D structure of MCs with distances up to 2.5 kpc (C. Zucker et al. 2021; T. E. Dharmawardena et al. 2023). These methods include using Gaia astrometric measurements coupled with dust extinction (M. M. Foley et al. 2023) or 3D position–position–velocity (PPV) maps using the radial velocity of CO combined with dust mapping techniques (C. Zucker et al. 2021; T. E. Dharmawardena et al. 2023). In this paper, we introduce a method to map out individual MCs in 3D using X-ray echoes that propagate throughout the CMZ and physically interact with MCs.

The X-ray emission from the Galactic center includes a soft plasma around 1 kT ~ 1 keV (G. Ponti et al. 2015), X-ray emission through point sources (e.g., Q. D. Wang et al. 2002; M. P. Muno et al. 2003, 2009; Z. Zhu et al. 2018), extended but not diffuse features (e.g., S. Zhang et al. 2020; E. Churazov et al. 2024), a hot and diffuse emission revealed by 6.7 keV Fe XXV (e.g., S. Park et al. 2004; K. Koyama 2018; K. Anastasopoulou et al. 2023), and a nonthermal component including a strong 6.4 keV Fe Kα line that correlates with molecular gas (e.g., R. A. Sunyaev et al. 1993; K. Koyama et al. 1996; H. Murakami et al. 2000; R. Terrier et al. 2018). The energetics, time variability, spectral shape, and polarization properties of the latter component are all compatible with the Fe Kα line being due to the past X-ray emission of Sgr A* (e.g., G. Ponti et al. 2010; R. Terrier et al. 2010; M. Clavel et al. 2013; R. Terrier et al. 2018; F. Marin et al. 2023). This past signal is propagating away from the supermassive black hole and interacts with MCs, creating a continuum emission due to X-ray scattering, which is absorbed at low energy. It also creates fluorescent emission lines, including the Fe Kα line that dominates the spectrum (R. Sunyaev & E. Churazov 1998). Furthermore, the flux of this fluorescent line is proportional to both the luminosity of the past event and the column density of the illuminated material. The X-ray monitoring of these clouds over the past ~25 yr (e.g., Y. Sofue 2000; M. Clavel et al. 2013; E. Churazov et al. 2017; D. Chuard et al. 2018; G. Ponti et al. 2019; P. Predehl et al. 2020; F. Marin et al. 2023) provides a unique opportunity to construct the 3D structure of MCs in the CMZ using the time lags of each X-ray observation as the third axis.

This paper is the second in a series to model the 3D structure of MCs in the CMZ using X-ray tomography. S. Brunker et al. (2025) introduced this method for the Sticks cloud located at $(\ell, b) = (0°.105, -0°.080)$. In this paper, we focus on the adjacent Stone cloud located at $(\ell, b) = (0°.068, -0°.076)$ as seen in Figure 1. The Stone cloud is $3.0 \times 10^4$ $M_\odot$ with a radius of 1.9 pc (C. Battersby et al. 2024a) and has a star formation rate of $(2.2 \pm 1.3) \times 10^{-10}$ $M_\odot$ yr$^{-1}$ (H. P. Hatchfield et al. 2024). Similar to the Sticks MC, the Stone MC has been observed by X-ray observatories such as Chandra and XMM-Newton and has been studied in recent literature (e.g., G. Ponti et al. 2010; R. Capelli et al. 2012; M. Clavel et al. 2013; K. Koyama 2018).

In this study we use X-ray emission from the Stone MC to construct a 3D visualization of the MC while characterizing its 3D structure. Section 2 gives an overview of the data used in this study, while Section 3 describes the methods used to disentangle the 3D structure of the Stone cloud and estimate its density profile by comparing with continuum and spectral line submillimeter emission. In Section 4 we provide a discussion on the new insights gained from this method and the shortcomings that it may possess. In Section 5 we summarize the key takeaways of this method.

## 2. Data

### 2.1. X-Ray Echoes

The X-ray observations of the Stone cloud were obtained using archival ACIS-I data from the Chandra X-ray Observatory covering the period 1999–2017. The data were reduced using CIAO v4.8 and following the method described in M. Clavel et al. (2013), to obtain maps of the continuum-subtracted 6.4 keV emission line.[4] To achieve a higher signal-to-noise ratio, the data are merged into one mosaic per year, except for 2012 and 2014, where no ACIS-I observation is available. These Fe K$\alpha$ mosaics tracing the X-ray echoes propagating in the Stone cloud have deep exposures in 2002, 2004, 2008, 2009, 2011, 2015, 2016, and 2017, with total clean exposures above 100 ks and shallower ones ranging from 25 ks (in 2003) to 80–90 ks (in 2010 and 2013). Subsequent observations show a decreasing trend (I. Khabibullin et al. 2022). Therefore, only the years 2008–2017 are used in the subsequent analysis of this paper.

### 2.2. Hi-GAL Column Density Map

The CMZ was observed with the Herschel Space Telescope through the Herschel infrared Galactic Plane (Hi-GAL) survey (S. Molinari et al. 2010, 2016) covering the Galactic plane at $|\ell| < 70°$ and $|b| < 1°$. The Spectral and Photometric Imaging Receiver (M. J. Griffin et al. 2010) and the Photodetector Array Camera and Spectrometer (A. Poglitsch et al. 2010) on Herschel observed the Galactic center at wavelengths of 70, 160, 250, 350, and 500 $\mu$m with beam sizes of 6″, 12″, 18″, 25″, and 36″ respectively. The column densities of MCs in this survey were derived by subtracting the cirrus emission from the Galactic center and performing modified blackbody fits to the cold dust component of the Herschel data. We refer the reader to C. Battersby et al. (2011, 2024b) for a complete explanation of this method.

### 2.3. CMZoom Survey

The CMZoom survey (C. Battersby et al. 2020) is a large-scale survey conducted with the Submillimeter Array (SMA) over 550 hr covering 350 arcmin² of the Milky Way's CMZ. The CMZoom survey mapped 1.3 mm continuum and spectral line emission in the CMZ above a molecular hydrogen column density of $10^{23}$ cm⁻². At these wavelengths we are most sensitive to cold, dense, star-forming gas and dust. These data have a resolution of 3″, or 0.1 pc, at a Galactic center distance of 8.2 kpc (GRAVITY Collaboration et al. 2019), allowing us to study MCs in great detail. The dust continuum map for the Stone MC can be found in Figure 1.

In addition to a continuum map, we use several molecular tracers to map the Stone cloud. These include $H_2CO$ 3(0,3)–2 (0,2) at 218.2 GHz and $H_2CO$ 3(2,2)–2(2,1) at 218.5 GHz, which map the dense gas, while the SiO (5–4) transition at 217.1 GHz is used to map protostellar outflows and shocks (D. Callanan et al. 2023). We use molecular tracers from the CMZoom survey because of its high resolution, revealing structures at subparsec scales. We use $H_2CO$ and SiO because they are well-known tracers of dense gas. We acknowledge that the choice to use only these two molecular lines only gives us a partial view of the cloud, particularly gas that is dense and hot.

### 2.4. Point Process Mapping

We compare with column densities from K. A. Marsh et al. (2017), which uses the PPMAP technique (K. A. Marsh et al. 2015). These column densities are based on Hi-GAL continuum maps from 70 to 500 $\mu$m and use the PPMAP to do a point process technique to perform a multicomponent fitting of the data. The PPMAP procedure provides improved resolution (12″) over other techniques. For more details, please see K. A. Marsh et al. (2017).

## 3. Methods and Results

### 3.1. Assumptions Based on the Geometry of X-Ray Echoes

The X-ray emission of the Stone MC has been interpreted as an echo from a past outburst from Sgr A* that occurred before the advent of X-ray astronomy and lasted at most 1 or 2 yr (G. Ponti et al. 2010; M. Clavel et al. 2013; E. Churazov et al. 2017). In this context, at a given time $t$ after the event, the material seen as illuminated by an observer on Earth follows the paraboloid

$$z(t) = \frac{1}{2}\left(ct - \frac{d_p^2}{ct}\right), \qquad (1)$$

where $c$ is the speed of light, $z(t)$ is the line-of-sight distance of the illuminated material, and $d_p$ is the distance between Sgr A* and the MC in projection (R. Sunyaev & E. Churazov 1998). The time in which light travels to the observer when interacting with an MC will include a time delay for the specific distance the light travels from Sgr A* to the MC ($d_p$). To take into account the time delay, we can write the total time in which light travels as $\left(ct - \frac{d_p^2}{ct}\right)$. In addition, the factor of $\frac{1}{2}$ represents the two separate paths light takes to reach the observer, i.e., light travels from the source to the material and then from the material to the observer. Therefore, Equation (1) can be used to link the 3D geometry of the X-ray emission to the age of Sgr A*'s past event, which is, however, poorly constrained. Recent estimations from various regions within the CMZ gave ages ranging from one to a few centuries (M. Clavel et al. 2013; E. Churazov et al. 2017; D. Chuard et al. 2018; F. Marin et al. 2023). Due to projection effects, the distance between two parabolas separated by 1 yr is larger than a light-year for an MC having a negative line-of-sight distance and smaller than a light-year for an MC having a positive line-of-sight distance.

---

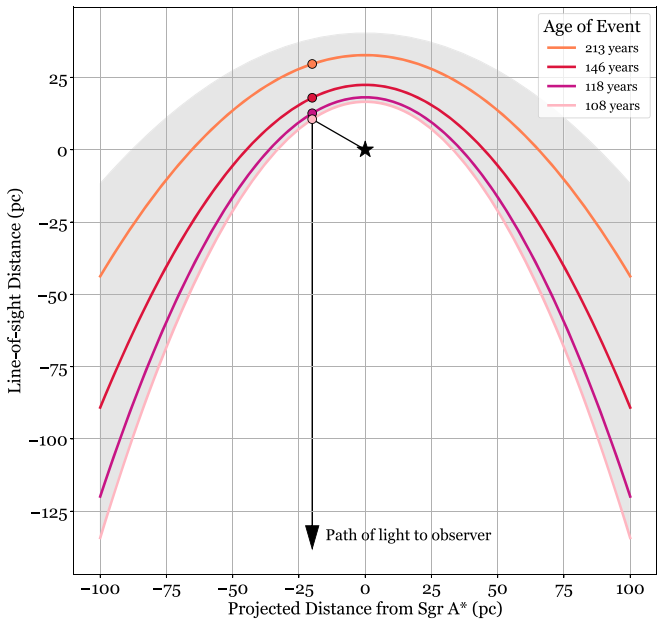

**Figure 2.** Top-down view of the CMZ with Sgr A* plotted at (0, 0) and parabolas following Equation (1). We consider scenarios where the X-ray flare propagated 108, 118, 146, and 213 yr ago (M. Clavel et al. 2013; E. Churazov et al. 2017; D. Chuard et al. 2018; F. Marin et al. 2023; as of 2008 January 1) with errors (shaded gray area). The circles correspond to the line-of-sight distance of the Stone MC given the different ages of the events.

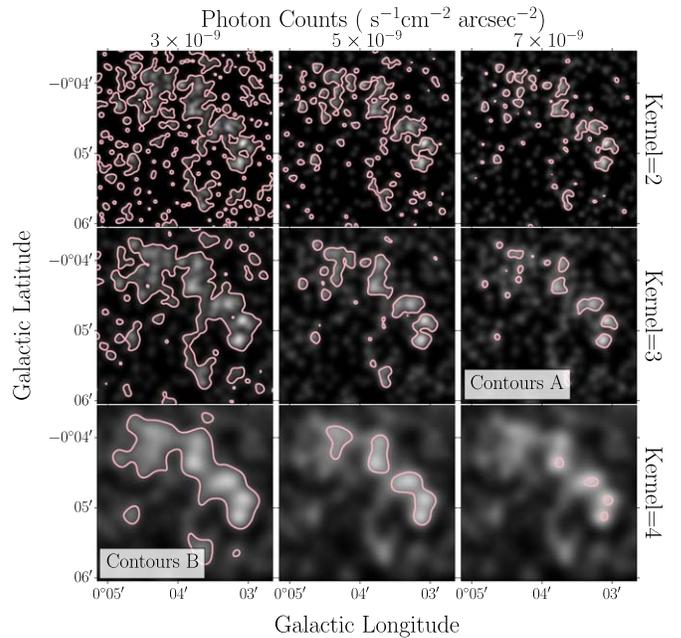

**Figure 3.** We investigate possible X-ray smoothing kernels and contour levels in order to select the ones that best represented the overall structure seen in X-rays. This figure shows X-ray data of the Stone (left) MCs collected in 2010. Shown from top to bottom are increasing numbers of kernels used for smoothing. From left to right, increasing flux levels of contours are drawn. In this work we use the middle right and bottom left panels (Contours A and B, respectively) to show the overall structure and the densest parts of the Stone MC.

In this paper we assume an age for the X-ray flare by using the calculated ages from previous studies (M. Clavel et al. 2013; E. Churazov et al. 2017; D. Chuard et al. 2018; F. Marin et al. 2023). Figure 2 shows all the possible scenarios for the X-ray flare event given different age estimates using 2008 as the reference year, with errors stemming from the literature values. The median of this sample is 132 yr, while the mean is 140 yr with a standard deviation of 39.72 yr. Although the median is often used to represent the average of a sample, we use the mean value for this work because of the small sample size available. Assuming that the age of the X-ray flare event was 140 yr ago, the line-of-sight distance can be calculated to ∼17$^{+21}_{-6}$ pc behind Sgr A* by using Equation (1). Errors for the line-of-sight calculation come from the errors in previous works, which are also shown in Figure 2. Other studies suggest that the Stone MC (commonly referred to as a section or multiple sections in the Bridge region) is 16.3–21.6 pc (R. Capelli et al. 2012), ∼10 pc (E. Churazov et al. 2017), and 18 pc (G. Ponti et al. 2010) behind Sgr A*. All line-of-sight distance estimates in previous works are based on different assumptions in the age of the event and definitions of this region.

### 3.2. The Structure of the Stone MC as Seen by X-Ray Echoes

Chandra X-ray Observatory, like other X-ray observatories, measures X-rays per photon on their detector, along with their energies and spatial information. Noise in an image can be introduced owing to the natural low photon count rates, stochastic nature of X-ray emission, and noise introduced when the photons hit the detector. These random fluctuations from an object can create a grainy background, concealing the origin of the X-ray emission. To make sure we are isolating the X-ray emission from the MC, we minimize background noise in the continuum-subtracted 6.4 keV maps by smoothing yearly mosaics using a Gaussian smoothing kernel. Averaging the pixel values of the photon counts within neighboring pixels attenuates noise, increasing the signal-to-noise ratio. However, there is a delicate balance between smoothing just enough to minimize noise and reveal astronomical objects and smoothing too much, which can remove important smaller features.

Figure 3 shows different smoothing kernels considered for the X-ray data, with different levels of flux drawn as contours. If we use a higher smoothing kernel, less noise is introduced, leading to a higher signal-to-noise ratio; however, the resolution of the data decreases.

To show the overall structure with minimum noise and high resolved areas, we use two combinations of smoothing and flux levels (see Figures 3 and 4), which are used in Paper I of this series (S. Brunker et al. 2025). We use a Gaussian smoothing kernel of 4 and a contour level of $3 \times 10^{-9}$ photons cm$^{-2}$ s$^{-1}$ arcsec$^{-2}$ to trace the overall shape of the Sticks MC and a smoothing kernel of 3 and contour level of $7 \times 10^{-9}$ photons cm$^{-2}$ s$^{-1}$ arcsec$^{-2}$ (hereafter referred to as Contours A and B, respectively) to identify highly resolved high flux areas. We used the lowest exposure year (in this case 2010) to define the contours that separate true features from background fluctuations. Therefore, everything that is above the contour threshold is a reliable structure, and what is below is noise. The rms values for each year are noted in Figure 4. Furthermore, given the density estimation for the Stone MC, we can safely assume that X-rays >7.1 keV producing the reflection features at 6.4 keV will propagate through the entire cloud, creating accurate slices throughout the MC. To have a consistent comparison between the Stone MC, the Sticks MC, and any other MC we choose to construct moving forward, we will use the same contour levels and smoothing kernel. We see that the X-ray emission is prominent in years 2008–2013, with smaller pockets of higher X-ray emission embedded in the thicker contours. However, after 2013 the MC lacks a clear structure and the higher X-ray emission is lost, indicating that the light front has left the densest part of the cloud.

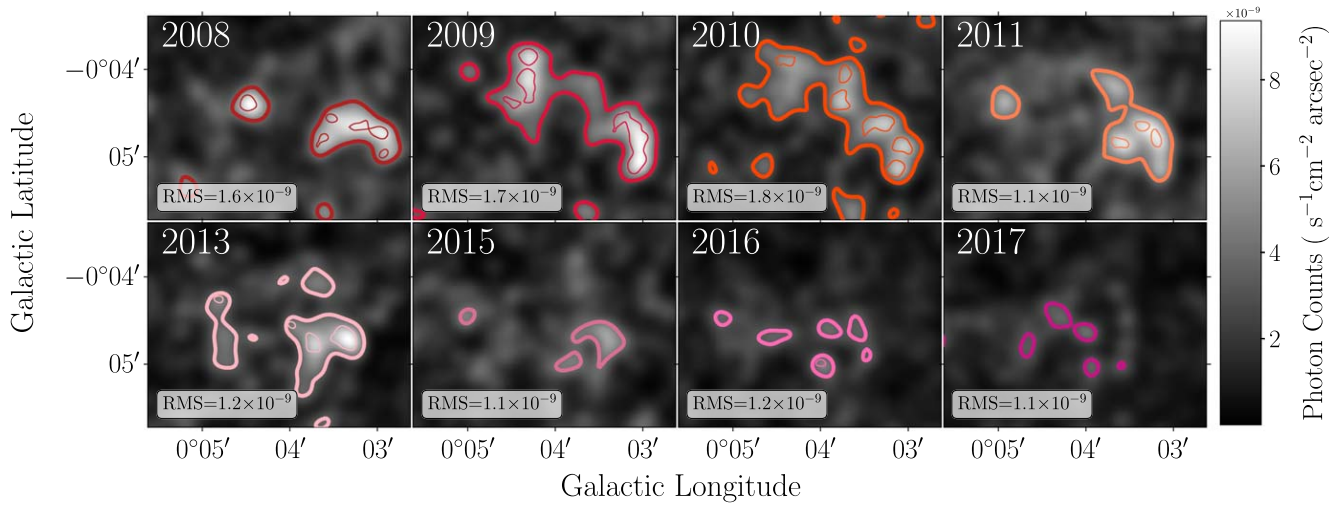

**Figure 4.** X-ray data collected from 2008 to 2017 show the most significant 6.4 keV emission in the vicinity of the cloud. The X-ray emission is plotted in gray scale, smoothed with a Gaussian kernel of 4, with rms values for each year in the lower left corner. Overplotted are Contours A in thick lines and Contours B in thin lines.

In order to create 3D models of MCs, we use the X-ray contours drawn in Figure 4 and the assumptions made in Section 3.1 to convert the time lags to physical distance. We then use the physical distance as the third axis in creating the Stone cloud's 3D structure (see Figure 5). We can see that there is a cohesive structure between the years 2008 and 2013, while later years show smaller areas of X-ray emission that don't match the pattern seen in earlier years. Furthermore, we can estimate the lower limit of the extent to be $1.66^{+0.18}_{-0.20}$ pc based on the standard deviations of the age of the event and Equation (1) described in Section 3.1.

### 3.3. Spectral Line Dendrograms

We use a hierarchical structure algorithm, astrodendro,[5] to identify structures within the $H_2CO$ 3(0, 3)–2(0, 2) at 218.2 GHz for the Stone MC. A dendrogram consists of trunks, i.e., structures with no parent structures and with the lowest emission, and branches, which split into multiple substructures called leaves. The leaves of a dendrogram have no additional substructure to them and are thus the highest level of structure with the brightest emission.

Specification of a dendrogram requires three parameters: the minimum value considered to be a structure, minimum delta or the difference of the peak flux of one structure and the structure that branches from it, and minimum number of pixels that constitutes a structure as a separate entity.

The dendrogram algorithm is used to tease out the structures located within the Stone MC of the $H_2CO$ (218.2 GHz) data cube where each channel has a width of 1.1 km s$^{-1}$. For each velocity spanning from 30.2 to 73.8 km s$^{-1}$ (40 channels in total) the median and standard deviation of the background are calculated. Furthermore, we calculate the rms values for the data by picking 10 random circles in the background of every quarter channel and finding the average rms value. The average rms values of the 73.8, 59.28, 44.7, and 30.2 km s$^{-1}$ channels are 0.05, 0.05, 0.05, and 0.04 Jy beam$^{-1}$, respectively. Therefore, the average rms value for all 40 channels is 0.05 Jy beam$^{-1}$. The minimum delta value is $3\sigma$ above the mean value, while the minimum delta parameter is set to $8\sigma$ above the mean value. The minimum number of pixels was set to $3'' \times 3''$, which is equivalent to the beam size of the CMZoom survey. For more information about the beam sizes for different regions of the CMZ, see C. Battersby et al. (2020).

The algorithm finds two trunks, 19 branches, and 21 leaves. The regions with the highest flux, or the leaves, of the dendrogram can be seen in Figures 6 and 7. Each leaf structure has a corresponding index number for identification that is labeled in Figure 6 and shown in Table 1 (except leaf 10, which is both a trunk and a leaf, and leaf 24 because its central velocity is an outlier). The index IDs of each structure in the dendrogram have no physical meaning but are randomly assigned by the algorithm. For each of the leaves in Table 1 the $\Delta v$ is identified as the total velocity extent of each structure and the $\bar{v}$ are the mean values of the total velocity extent. The leaves are grouped together from 30.2 to 40 km s$^{-1}$, from 40 to 50 km s$^{-1}$, and from 50 to 63.8 km s$^{-1}$ based on the mean central velocities of each in Table 1 and are shown in Figure 6. In addition, the X-ray years are also split into these three groups based on their mean $\bar{v}$, which are calculated by calculating the mean $\bar{v}$ of all the structures that correspond to it. If multiple years are grouped within the same velocity range, the outlines of their corresponding contours (Contours A) are added together.

We then match each leaf with an X-ray year by overlaying the structures found in each velocity and comparing their positions to the area covered by the X-ray echoes (see Figure 7). In addition, we use the central velocities found in Table 1 to further constrain the groupings. For example, if a leaf is in the same projected position as two separate X-ray years, we favor the year in which other leaves have similar velocities. This way we are able to group structures together in PPV space and their projected area. All leaves except 21 and 25, which are shown in gray scale in Figure 7, are given a corresponding X-ray year.

The regions with lower flux, trunks and branches, can be seen in Figure 8, where each structure is integrated from the velocity extent of the MC (30.2–73.8 km s$^{-1}$). We refer the reader to Section 3.4 for an in-depth analysis.

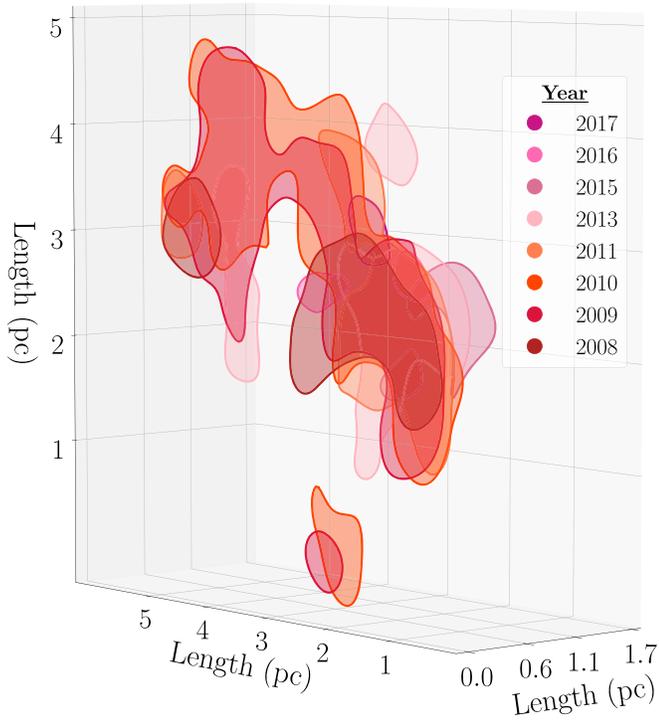

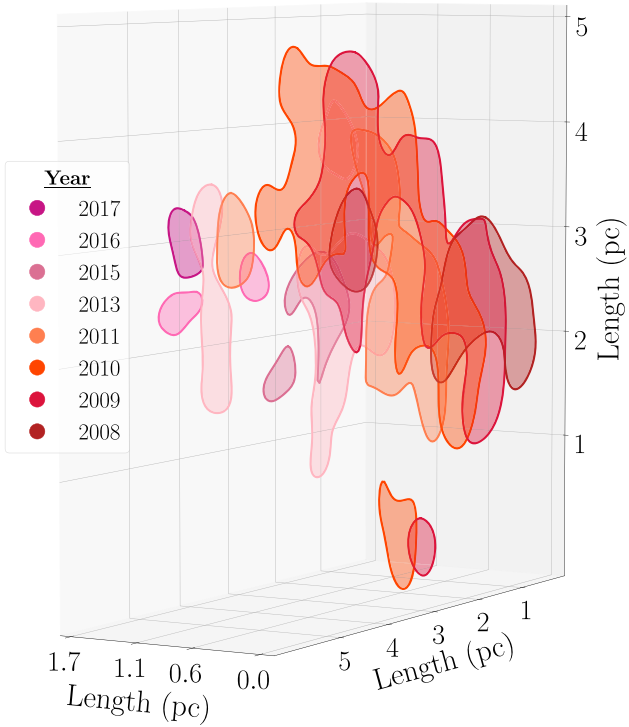

**Figure 5.** Using Equation (1) and assuming an X-ray flare event ∼140.6 yr ago (See Section 3.1) where $d_{los} = 17$ pc, we are able to create a 3D model of the Stone MC with Contours A. Structures that were less than 15″ wide were omitted from the plot to omit background fluctuations.

### 3.4. Integrated Structure Maps

We transform 3D information in the form of X-ray echoes and spectral line data cubes into 2D images to compare them to data sets that only observe 2D structure (such as column densities).

First, we use two different methods to create integrated X-ray contours shown in Figure 8. Integrated X-ray contours are created emphasizing the following:

1. Maximum X-ray brightness per year where lower-level contours (Contour A) shown in Figures 4 and 5 (with contours less than 15″ omitted) are used. The integrated contour is then drawn from the outline of the individual years added together. This gives more weight to clumps in the MC, as they are localized areas that are less dense than the overall MC but still pass the minimum threshold flux.

2. Relative X-ray brightness over all years where the X-ray data are smoothed with a kernel of 4 and all the years between 2008 and 2017 are added together. The integrated contour is then drawn based on this map at $1.2 \times 10^{-8}$ photons cm$^{-2}$ s$^{-1}$ arcsec$^{-2}$. Unlike the first method, this second method gives more weight to extended features in the X-ray data.

Additionally, we produce an integrated image with additional spectral lines of different molecules. We use $H_2CO$ 3(0,3)–2(0,2), $H_2CO$ 3(2,2)–2(2,1), and SiO (5–4) to trace the dense molecular gas in the Stone MC. Each is integrated over the velocity range of the cloud (30.2–73.8 km s$^{-1}$) and assigned a color, which is then overlaid in a single plot (see Figure 9).

Figure 8 compares the integrated X-ray contours with other data seen in the line of sight, including the CMZoom continuum and Herschel Hi-GAL column densities and the branches and leaves from the dendrogram. We see that the maximum integrated contour does not trace the densest structures but instead encompasses the less dense areas of the MC. Conversely, the relative integrated contour is able to trace the outer arc and the majority of the inner arc but fails to trace a small part of the brightest point (top left circular area) of the MC. Meanwhile, both X-ray integrated contours fail to trace over the lower center structures (the leftmost parts of the inner arc).

On the other hand, Figure 9 shows that all three spectral lines are in agreement with each other and the structure of the Stone MC seen in the submillimeter continuum (see Figure 1), showing that it is a reliable tracer of dense gas in the Stone MC.

### 3.5. Calculating Localized Densities

For further analysis of the 3D properties of the Stone MC we calculate the density of material for each relevant year. We make the assumption that the relationship between the X-ray emission and Herschel column densities is linearly proportional; however, the positions of the peak flux in the relative X-ray brightness integrated maps (as described in Section 3.4) and that in the Herschel column density differ by about 30″ (for reference, the beam size for Herschel 350 $\mu$m emission is 35″). Since the X-ray echoes are not measured continuously, this could indicate that the brightest parts of the MC are missed by Chandra. Although this is an issue for our assumption, we move forward with the calculation, hoping that future studies can improve this calculation.

In order to create localized densities for each year of the cloud, we identify the peak X-ray intensity of the integrated X-ray flux (see Figure 8) and the peak column density from

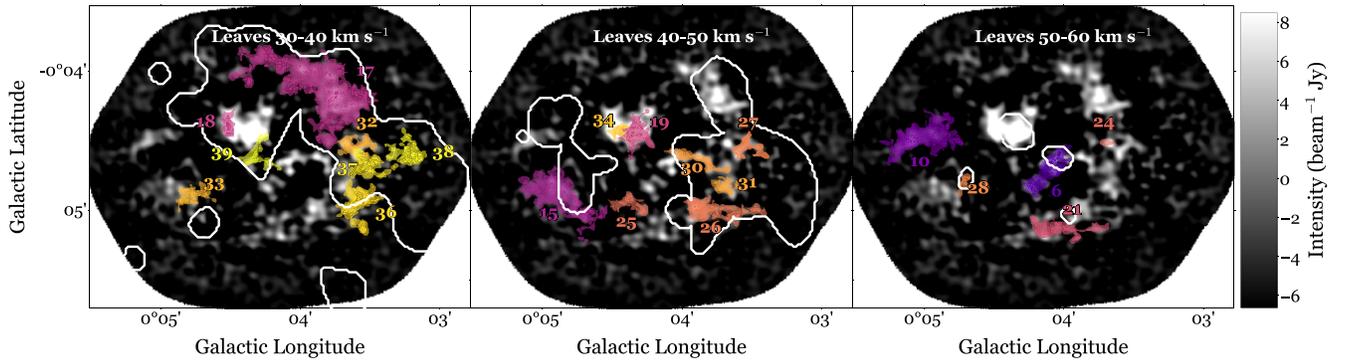

**Figure 6.** Grouping structures with similar velocity ranges reveals an overall trend in the leaves seen in PPV space. Left panel: leaf indices 17, 18, 32, 33, 36, 37, 38, and 39. Middle panel: leaf indices 15, 19, 25, 26, 27, 30, 31, 34. Right panel: leaf indices 6, 10, 21, 24, and 28. Integrated X-ray contours are created based on the mean $\bar{v}$ in Table 1 and shown as white contours in each panel (left: 2008, 2009, 2010; middle: 2011, 2013, 2015, 2016; right: 2017). Leaves 21 and 25 are not matched to any X-ray years, while leaves 10 and 24 are not included in the analysis but are shown in case the reader is curious about their central velocities and projected positions.

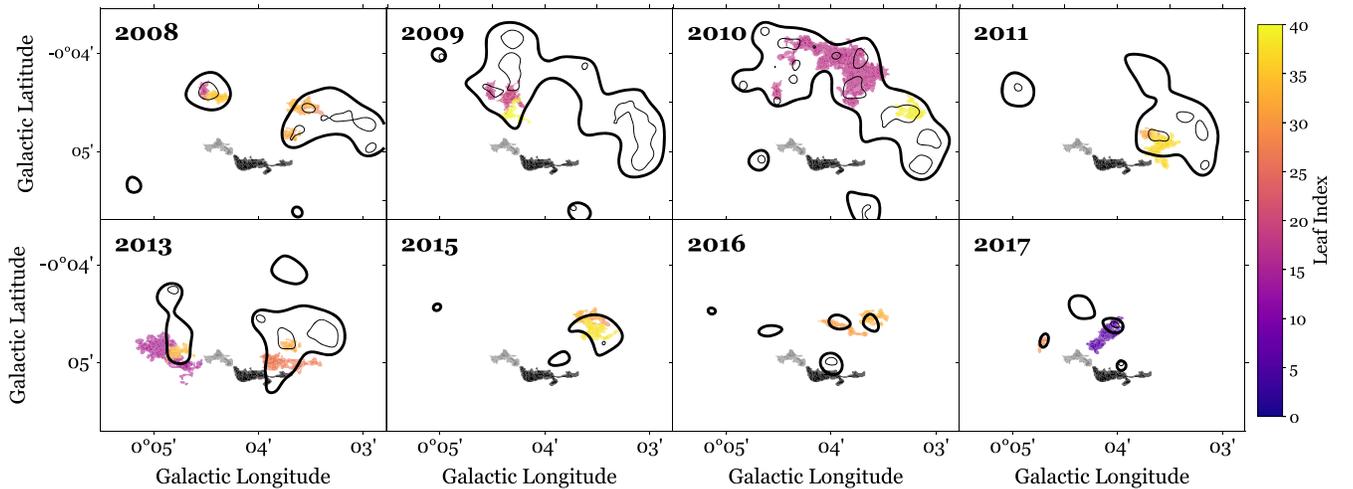

**Figure 7.** Dendrogram leaves of the $H_2CO$ spectral line show good correspondence with X-ray contours. Structures colored from yellow to purple represent different structures that are correlated, while the structures in gray scale are unmatched with a particular feature in the X-ray. All dendrogram structures are seen in the $H_2CO$ spectral line within 30.2–73.8 km s$^{-1}$. Contours A are drawn with thick lines, while Contours B are drawn with thin lines. For detailed information about the overlaid data sets and the velocity ranges for each dendrogram structure, see Table 1.

Herschel. We then divide the peak flux in the column density by the peak X-ray flux to find a density normalization factor of $1.23 \times 10^{31}$ s arcsec$^{-2}$. This normalization factor is then multiplied by the X-ray flux in each observation to obtain a localized density for every pixel in each X-ray year. Figure 10 shows that years 2008, 2009, and 2010 contribute the highest densities at around $1 \times 10^{23}$ cm$^{-3}$. Moreover, there are smaller dense regions illuminated in 2009 and 2010, which then become less dense in 2011 and 2013, disappearing after 2013. Since the CMZoom continuum integrates all of the dust along the line of sight while the X-ray emission only covers the emission observed at a certain time, the continuum is a fixed number and the X-ray emission is a lower bound. If we were to observe more X-ray emission, then the normalization factor would only decrease, meaning that our density calculation for each year is an upper limit.

## 4. Discussion

### 4.1. The Spectral Line Counterparts of the X-Ray Echoes

Typically, Doppler velocities cannot be assumed to follow linear time such as the X-ray flux, especially if there are other factors such as turbulence or star formation that could skew the velocity ranges detected. By following each substructure through different velocities, we show that most of the lower-emission X-ray contours (Contours A) cover the leaves from the dendrogram. While some of these leaves are completely covered by the X-ray emission, other leaves are only partially covered, while very few (about two leaves) are not covered by the X-ray data at all. The velocity dispersion for each leaf in Table 1 shows a wide range from 15.7 to as low as 1.2 km s$^{-1}$. This means that some structures in the Stone MC are extended, while structures with small velocity dispersions can be described as small clumps of dense gas. In addition, we find average velocities for each year, along with their standard deviation and error. The standard deviation shows a small spread of velocities ($\lesssim$5 km s$^{-1}$) for all years. Indices 27, 18, 31, and 32 have very wide velocity spreads (12.3, 14.6, 10.0, and 15.7 km s$^{-1}$, respectively).

In addition, the branches and trunks of the dendrogram are compared with the X-ray integrated contours in Figure 8. Here we can see that both integrated contours match the overall shape of the dendrogram except in the middle, where the top part of the inner arc is not covered by X-rays. On the far right area, this mismatch can be attributed to the limited field of view of the CMZoom survey, which is less sensitive to emission in

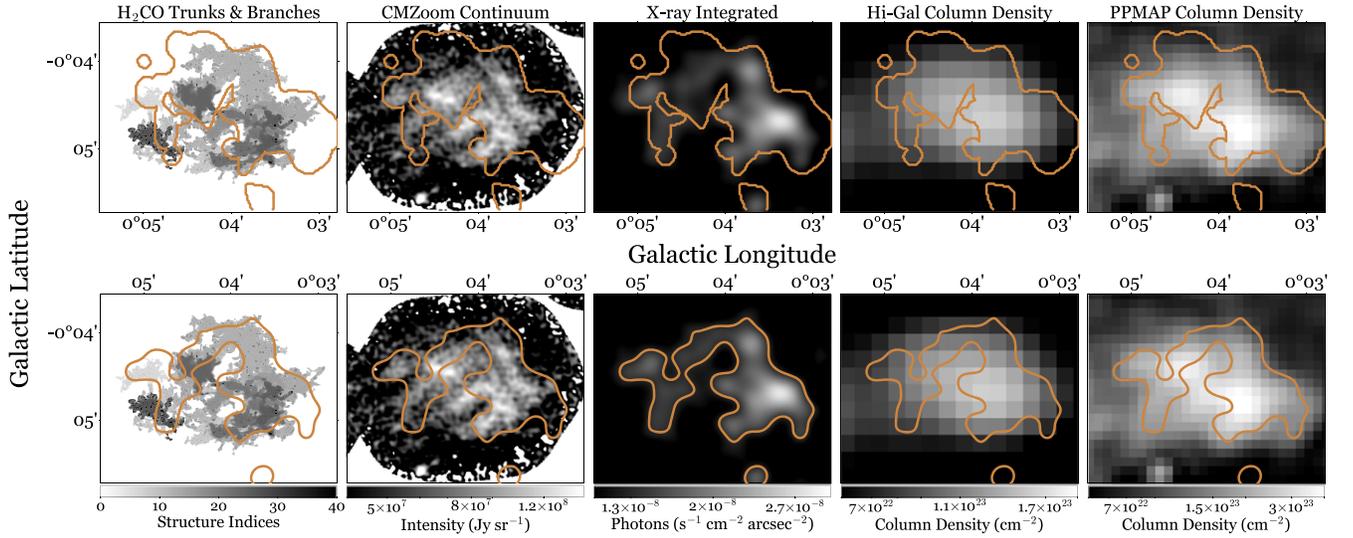

**Figure 8.** Integrated X-ray contours highlight the overall structure of the Stone MC. Top row: maximum X-ray brightness contour; bottom row: relative X-ray brightness contour (see Section 3.4 for more details). From left to right: branches and leaves of the H$_2$CO (218.2 GHz) dendrogram colored by structure ID, CMZoom 1.3 mm continuum, X-ray 6.4 keV emission integrated from 2008–2017, Hi-GAL column density, and PPMAP column density.

**Table 1**
Leaf Structures in the Dendrogram Matched with the X-Ray Year Contours, Showing That the Structures in the Same Year Are in Similar Areas in PPV Space

| Year | Index | $\Delta v$ (km s$^{-1}$) | $\bar{v}$ (km s$^{-1}$) | Mean $\bar{v}$ (km s$^{-1}$) | Standard Deviation | Standard Error |
|------|-------|------|------|------|------|------|
| **2008** | 27 | 12.3 | 45.4 | 40.0 | 3.5 | 1.4 |
| | 18 | 14.6 | 37.5 | ... | ... | |
| | 31 | 10.0 | 43.1 | ... | ... | |
| | 32 | 15.7 | 38.1 | ... | ... | |
| | 34 | 5.5 | 40.9 | ... | ... | |
| **2009** | 19 | 5.4 | 42.1 | 38.5 | 2.6 | 1.5 |
| | 18 | 14.6 | 37.5 | ... | ... | |
| | 39 | 8.9 | 35.9 | ... | ... | |
| **2010** | 17 | 15.7 | 38.1 | 37.0 | 1.2 | 0.7 |
| | 18 | 14.6 | 37.5 | ... | ... | |
| | 38 | 10.1 | 35.3 | ... | ... | |
| **2011** | 31 | 10.0 | 43.1 | 40.6 | 2.5 | 1.7 |
| | 36 | 15.7 | 38.1 | ... | ... | |
| **2013** | 26 | 10.1 | 48.7 | 43.8 | 3.9 | 2.0 |
| | 15 | 12.3 | 45.4 | ... | ... | |
| | 31 | 10.0 | 43.1 | ... | ... | |
| | 33 | 13.4 | 38.1 | ... | ... | |
| **2015** | 27 | 12.3 | 45.4 | 40.3 | 3.6 | 2.0 |
| | 32 | 15.7 | 38.1 | ... | ... | |
| | 37 | 14.6 | 37.5 | ... | ... | |
| **2016** | 30 | 3.3 | 47.6 | 42.8 | 4.75 | 3.4 |
| | 32 | 15.7 | 38.1 | ... | ... | |
| **2017** | 28 | 3.3 | 58.8 | 59.7 | 0.8 | 0.6 |
| | 6 | 13.3 | 57.2 | ... | ... | |
| **Other** | 21 | 13.4 | 57.1 | 52.5 | 4.6 | 3.2 |
| | 25 | 15.9 | 48.0 | ... | ... | |

**Note.** There are only two leaves that do not correspond to an X-ray contour (see the "Other" section at bottom). Leaves considered but not included with the groupings are Index 10 because it is both a leaf and a trunk and Index 24 because of its large average velocity. We include these leaves in Figures 6 and 7 in case the readers are interested in identifying and seeing these structures. From left to right: X-ray years; the index of the corresponding dendrogram structure (only the leaves are used for this analysis); $\Delta v$, average velocity for each structure; and the average velocity of all structures for their corresponding year. All values in the table have rounding errors of $\pm 0.1$ km s$^{-1}$.

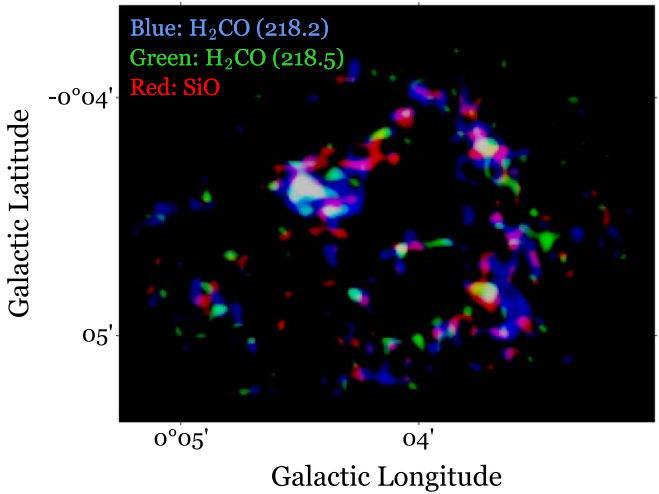

**Figure 9.** Spectral lines integrated reveal the 2D shape of the Stone MC from our line of sight. The spectral lines used were $H_2CO$ 3(0,3)–2(0,2), $H_2CO$ 3(2,2)–2(2,1), and SiO (5–4), plotted in blue, green, and red, respectively.

the outer areas of the circle. Moreover, when comparing both integrated X-ray contours with the Hi-GAL and PPMAP column densities that have a wider field of view, the X-ray contours encompass the cloud completely.

### 4.2. Do Certain Structures Occupy Similar Volumes in Space?

We analyze each leaf's location in PPV and projected space, to group leaves and X-ray echoes together. Interestingly, leaves in the 30.2–40 km s$^{-1}$ range are located in two separate areas: the larger arc (on the top and right side of the image) and small parts of the left side. Leaves in the 40– 50 km s$^{-1}$ range begin to fill in the right side of the smaller arc (near the middle of the image), the bright circle near the top left, and the outer left areas of the image. Lastly, the leaves in the 40– 63.8 km s$^{-1}$ range trace the leftmost part of the inner arc and the rest of the outer left structures in the image. Furthermore, when looking at the overall trend of all three panels, we see the structures appearing from right to left as the Doppler velocities increase. It is worth noting that leaf 10 (which can be seen in the right panel of Figure 6 as the purple leftmost structure) is a leaf and a branch, having no connection to the other leaves in the sample.

Similarly, while not perfect, the X-ray echoes have a similar trend. The emission that reaches the observer first (in 2008, 2009, 2010) outlines the larger arc, while the right side of the inner arc and the left side of the image are also highlighted in the years 2011, 2013, 2015, and 2016. Lastly, the structures on the left side of the inner arc and the left side of the image are congruent with the X-ray contours beginning to occupy the centermost and left sides of the Stone MC in 2017. However, about a third of the leaves in Figure 6 do not match the combined X-ray echoes that are overlaid in white. This alludes to the fact that although we see similar trends for both data sets, there is not a one-to-one comparison between them.

### 4.3. Coverage of the X-Ray Echoes

While we see very good overlap between the molecular line data and X-rays, there are some features in the molecular lines that are not seen in the X-rays (such as leaves 21 and 25), while leaf 24 is hard to correlate owing to its high central velocity and small projected area. We know from Section 3.3 that the majority of the leaves are clumped within similar velocity ranges and lie over the projected area of the cloud as shown in Figures 7 and 8. Thus, we can assume that these features are missing in the X-ray owing to the intermittent nature of X-ray observations. In addition, if we assume that the missing leaves are spherical, we can use the widths of the leaves to compute the upper limit of the duration of the X-ray flare event. We plot the area illuminated by the 140 yr old event (as of 2008 January 1) of different duration that would have been observed by the current data set (see Figure 11). For an infinitely short event, the extension of each parabola is driven by the duration of the observations, with larger spacing around 2012 and 2014. If the event lasts more than a few months, then the extension of the parabolas is driven by the duration of the event. The space available for possible missing clumps reduces and completely disappears for events longer than 1.5 yr. For an upper limit, where the event took place $t = 140$ yr ago, with a duration $= 0$ yr and $d_{proj} = -18$ pc, the maximum extension that could be missed in 2012 and 2014 is ~0.28 pc.

We estimate the extension along the line of sight to be equivalent to the diameter of the MC, i.e., ~0.3 pc. Therefore, the bulk of X-ray echoes from these clumps could have been missed by the Chandra coverage only if the illuminating event is not much longer than 5 months. Such a short-duration event is compatible with previous estimations based on X-ray variability studies of this MC (e.g., M. Clavel et al. 2013; E. Churazov et al. 2017).

### 4.4. Comparing the 3D Structure of the Sticks and Stone MCs

In Paper I of this series (S. Brunker et al. 2025), the Sticks MC was modeled using the methods outlined in Section 3 without the dendrogram analysis. Unlike the Stone MC, the features shown in Doppler-shifted spectral lines and the X-ray echoes progress similarly as velocity and time increases. Furthermore, for the Sticks MC the positions of the peak X-ray integrated flux and the peak column density matched. In contrast, we have no reason to believe that molecular line velocities would necessarily linearly correlate with the physical distance in the Stone MC. In a simple, slowly rotating cloud, the velocity may trace the distance along the line of sight (as it seems to in S. Brunker et al. 2025); however, in the presence of feedback from forming stars, localized turbulence, or other chaotic contribution to the velocity field, this assumption would break down. Dendrogram analysis can provide insight into how structures that are defined in PPV space can be associated with similar physical structures highlighted by the X-rays. Furthermore, the larger X-ray extent as shown by the increase in years of X-ray emission and the velocity range extent of the Stone MC indicate that the Stone MC may extend farther than the Sticks MC in our line of sight. In addition, the dendrogram analysis revealed its complex substructure by identifying its many clumps, which span a wide range of velocity dispersions.

The difference in physical properties of the Sticks and Stone MCs should also be taken into consideration. D. Callanan et al. (2023) have shown that the Stone MC has an abnormally large velocity range for an MC of its size in projection. In particular, factors other than distance can contribute to this range of velocities, such as star formation or turbulence within the MC. H. P. Hatchfield et al. (2024) estimate the star formation rate of the Stone MC as $(2.2 \pm 1.3) \times 10^{-10}$ $M_\odot$, while the Sticks cloud does not show signs of current star formation. Thus, a combination of outside factors affecting the spectral line

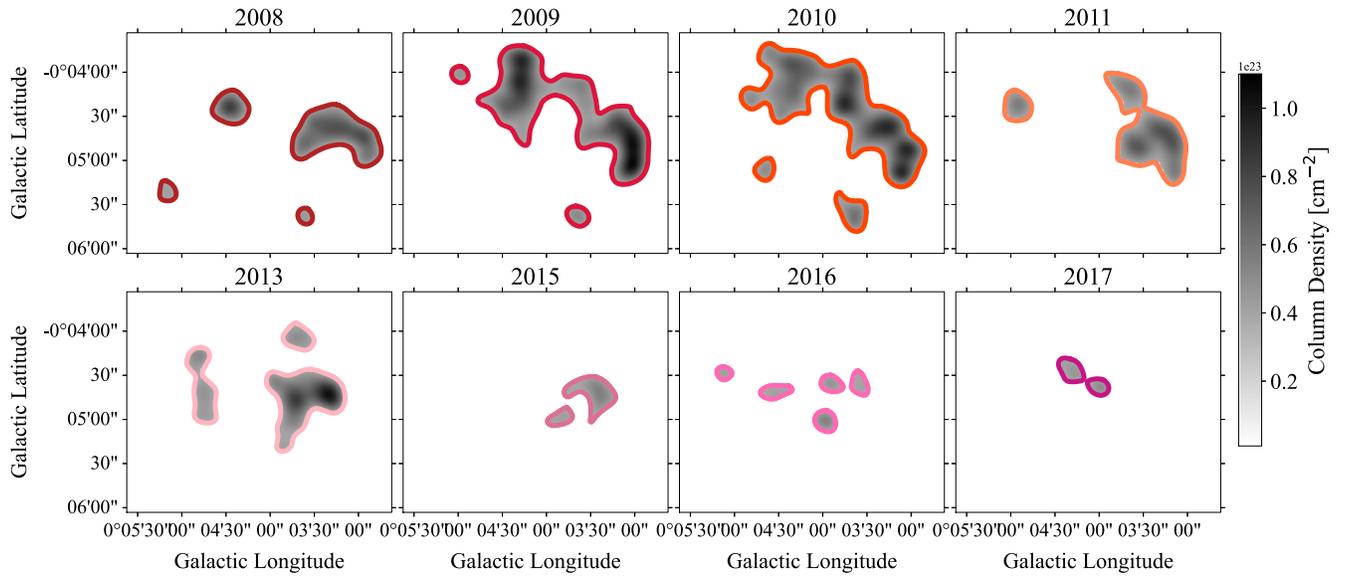

**Figure 10.** Assuming that the peak in X-ray emission correlates linearly with the peak of submillimeter continuum data from C. Battersby et al. (2020), we produce localized densities for each slice of the Stone MC. Contours are drawn at $3 \times 10^{-9}$ photons cm$^{-2}$ s$^{-1}$ arcsec$^{-2}$ from data smoothed with a kernel of 4.

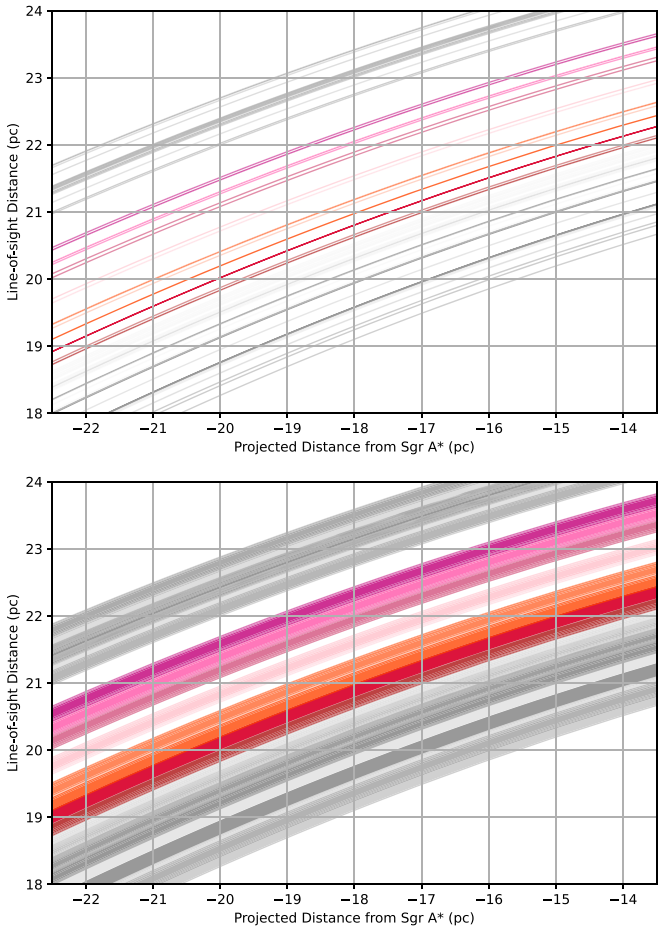

**Figure 11.** Line-of-sight coverage of existing Chandra observations at the projected distance of the Stone cloud, based on their starting date and duration. The parabolas follow Equation (1), assuming a 140 yr old event (as of 2008 January 1) with an infinitely short duration (top) and a 1 yr duration (bottom). The color indicates the observation year with the same color code as the X-ray contours, and existing observations not studied in this work are marked in gray. Finally, the shading gives an indication of the depth of the observations covering the plane.

Doppler velocities and the Stone MC's diffuse nature has made it more difficult to model than the Sticks MC.

### 4.5. Limitations

The X-ray tomography method uses several assumptions, and our results have limitations. We summarize some of these limitations in the list below.

1. The exact description provided by this method is dependent on the smoothing kernel and flux levels used for the contours. In this study we consider a wide variety of combinations in order to pick the parameters for our contours. Additionally, we create two separate contours that encompass the overall shape (higher-resolved/higher-flux areas) and densest parts (lower-resolved/lower-flux areas) of the MC. For more information see Section 2.1.

2. The distance between each year and the lower limit extent is computed using Equation (1), which depends on the age of the event—something that is still not well-known. However, in this particular paper, the errors are particularly low. For example, the errors attributed to the extent of the MC are ⩽0.20 pc. Furthermore, when creating the 3D model, the slices are along the X-ray line of sight from Sgr A* and not along the line of sight for us.

3. Since the X-ray observations are not continuous, there is a chance that the X-ray emission available does not completely trace all of the dense structures in the MC. For this reason, comparing the X-ray emission with data from other wavelengths is important to determine how accurate the X-ray echoes are in creating the 3D structure. This can have implications for the density calculations as well (see Section 3.5).

4. For this particular MC, we see the X-ray emission begin in 2008 and decrease substantially after 2017, indicating that the entire MC has been scanned. However, for other MCs this may not be the case. In this situation additional X-ray observations need to be taken in order to model the entire MC. Furthermore, you may also have big holes in existing data, which will make the analysis more complex owing to gaps in data or sparse time coverage.

5. The duration of the X-ray flare event is an assumption that impacts the way the 3D model is created. Since previous studies suggested a duration of $\lesssim 2$ yr, we are able to assume that a single year of X-ray observation is equal to one cross–sectional slice of the MC. However, if the X-ray flare event is longer, X-ray observations from each year will overlap, making this assumption incorrect. Figure 11 shows how the duration of the event can impact the coverage of the Stone cloud.

## 5. Conclusions

We have presented a new method using X-ray fluorescence to model MCs in the CMZ from X-ray emission observed over time. The emission is believed to propagate from the supermassive black hole, Sgr A*, after it accretes material. Over the period of $\sim 20$ yr, Chandra X-ray Observatory has detected variable emission from MCs in the CMZ, likely due to these X-ray echoes. We use the time lag between each X-ray fluorescence event of the Stone MC as the third axis in creating the 3D model. We have also compared the X-ray emission to various spectral lines from the CMZoom survey (D. Callanan et al. 2023) and created a dendrogram to analyze the distinct PPV structures in the MC and the hierarchical substructure of the MC. From this work, we perform the following:

1. We use Gaussian smoothing to improve the signal-to-noise ratio of each X-ray observation and specify two different combinations of kernels used for smoothing, along with a flux level used for contours. A higher smoothing kernel with a lower flux level is used to trace the overall shape of the MC, while a lower smoothing level with a higher flux is used to identify dense clumps. Furthermore, by assuming the age of the X-ray flare to be 140 yr and using Equation (1), we are able to use the time difference between each observation as a third axis, allowing us to create a 3D model of the Stone MC.

2. We identify substructures in the MC using `astrodendro`, a hierarchical structure algorithm, and compare (1) the highest-flux structures (leaves) with individual X-ray years and the (2) lowest-flux structures (trunks and branches) with the integrated X-ray flux. By identifying the central velocities of each leaf, we are able to match structures to specific X-ray years, combining the PPV areas and the projected areas of each leaf. Detecting X-ray emission that is correlated with the molecular gas indicates that we are seeing X-ray echoes moving through an MC and mapping it over time in 3D. Furthermore, the integrated X-ray emission covers most of the trunks and branches identified, missing an area of $\sim 0.3$ pc shown in the spectral lines and CMZoom continuum. On the other hand, the X-ray integrated emission covers the entire area of the MC in projection seen by Herschel, indicating that the missing X-ray data could be responsible for the structures in the $H_2CO$ data not covered by the X-rays.

3. By assuming a linear relationship between column densities and X-ray flux, we find the peak intensities of both data types and calculate their ratio. We use this ratio to compute a pixel-by-pixel column density map for each year of X-ray emission to find localized densities in our line of sight.

4. By assuming that the missing structures (leaves 21 and 25) seen in spectral lines are spherical clumps, we can constrain the duration of the X-ray flare event. We find that the projected length of the missing clumps is $\sim 0.3$ pc, meaning that the X-ray flare event illuminating the Stone MC could not have lasted more than 5 months.

5. From the 3D model of the Stone MC, we know that the line-of-sight extent of this MC is 1.7 pc while the projected length and width are about 5 pc. Furthermore, the molecular line structures indicate that the Stone MC has dense clumps shown by the leaves of the dendrogram that are surrounded by diffuse gas shown by the trunks and branches of the dendrogram.

We expect to model MCs using X-ray emission for about 5–10 additional MCs in the CMZ. The modeling of MCs in the CMZ helps us constrain their structure. However, much work still needs to be done in understanding them. For instance, the turbulence, star formation, and density distribution are still not well-known. The conclusions in this paper give preliminary insight into the MC's structure, and we hope that future studies, both observational and computationally, can further constrain these physical attributes for each MC. Furthermore, the small line-of-sight extent of the Stone MC is interesting, and further studies may shed insight on the extents for other MCs.

For Sgr A*, this analysis provides an upper limit on the duration of the X-ray flare event illuminating the Stone MC. By performing a similar analysis on other MCs in the CMZ emitting X-ray echoes, we will be able to constrain their durations as well. Constraining these durations provides insight into the frequency of X-ray flares, either supporting or refuting current X-ray flare models of the Sgr A*.

### Acknowledgments

D.A., C.B., and S.B. gratefully acknowledge funding from the National Aeronautics and Space Administration through the Astrophysics Data Analysis Program under award "3-D MC: Mapping Circumnuclear Molecular Clouds from X-ray to Radio," grant No. 80NSSC22K1125. Additionally, C.B. gratefully acknowledges funding from National Science Foundation under award Nos. 2108938, 2206510, and CAREER 2145689. D.A. acknowledges the Summer Undergraduate Research Fund Award and Research Travel Award from the Office of Undergraduate Research at the University of Connecticut, as well as the FAMOUS Travel Grant from the American Astronomical Society. M.C. acknowledges financial support from the Centre National d'Etudes Spatiales (CNES).

The scientific results reported in this article are based to a significant degree on observations made by the Chandra X-ray Observatory. This research also made use of data from the SMA, and the authors wish to recognize and acknowledge the very significant cultural role and reverence that the summit of Maunakea has always had within the indigenous Hawaiian community; we are most fortunate to have had the opportunity to utilize observations from this mountain.

*Software:* This research has made use of software provided by the Chandra X-ray Center (CXC) in the application package CIAO. This work also made use of Astropy,[6] a community-

---

developed core Python package and an ecosystem of tools and resources for astronomy (Astropy Collaboration et al. 2013, 2018; Astropy Collaboration et al. 2022); astrodendro, a Python package to compute dendrograms of Astronomical data;[7] SAO ds9 (W. A. Joye & E. Mandel 2003); and Matplotlib (J. D. Hunter 2007).

## ORCID iDs

Danya Alboslani  https://orcid.org/0009-0005-9578-2192
Cara Battersby  https://orcid.org/0000-0002-6073-9320
Samantha W. Brunker  https://orcid.org/0000-0001-6776-2550
Maïca Clavel  https://orcid.org/0000-0003-0724-2742
Dani Lipman  https://orcid.org/0000-0002-5776-9473
Daniel L. Walker  https://orcid.org/0000-0001-7330-8856

---

[7] http://www.dendrograms.org/